\documentclass[runningheads]{llncs}

\usepackage{graphicx}
\usepackage{algorithm} 
\usepackage{algorithmic}
\usepackage{multirow}
\usepackage{mathrsfs}
\usepackage{color}
\usepackage{diagbox} 
\usepackage{booktabs} 
\usepackage{amsmath}
\usepackage{amssymb}
\usepackage{marvosym}
\newcolumntype{M}[1]{>{\centering\arraybackslash}m{#1}}

\begin{document}
\title{Improving Information Cascade Modeling by Social Topology and Dual Role User Dependency
}
\author{Baichuan Liu$^{\S}$\orcidID{0000-0002-6216-3011}\and Deqing Yang$^{\S}$\thanks{Deqing Yang is the corresponding author. This work was supported by Shanghai Science and Technology Innovation Action Plan No.21511100400.} \orcidID{0000-0002-1390-3861} \and\\
 Yuchen Shi$^{\ddag}$ \and Yueyi Wang$^{\ddag}$}
\institute{ School of Data Science, Fudan University, Shanghai 200433, China. 
$^{\S}$\email{\{bcliu20,yangdeqing\}@fudan.edu.cn}, $^{\ddag}$\email{\{ycshi21,yueyiwang21\}@m.fudan.edu.cn}}
\titlerunning{Cascade Modeling by Topology and User Dependency}
\maketitle           

\begin{abstract}
In the last decade, information diffusion (also known as information cascade) on social networks has been massively investigated due to its application values in many fields. In recent years, many sequential models including those models based on recurrent neural networks have been broadly employed to predict information cascade. However, the user dependencies in a cascade sequence captured by sequential models are generally unidirectional and inconsistent with diffusion trees. For example, the true trigger of a successor may be a non-immediate predecessor rather than the immediate predecessor in the sequence. To capture user dependencies more sufficiently which are crucial to precise cascade modeling, we propose a non-sequential information cascade model named as  {\bf TAN-DRUD} ({\bf T}opology-aware {\bf A}ttention {\bf N}etworks with {\bf D}ual {\bf R}ole {\bf U}ser {\bf D}ependency). TAN-DRUD obtains satisfactory performance on information cascade modeling through capturing the dual role user dependencies of information sender and receiver, which is inspired by the classic communication theory. Furthermore, TAN-DRUD incorporates social topology into two-level attention networks for enhanced information diffusion prediction. Our extensive experiments on three cascade datasets demonstrate that our model is not only superior to the state-of-the-art cascade models, but also capable of exploiting topology information and inferring diffusion trees.

\end{abstract}

\section{Introduction}
With the development of online social networks, various information spreads more quickly and broadly on web. The diffusion of a piece of information generally forms a cascade among different users in the network, which is often observed as a sequence of \emph{activated users} who have disseminated the information. The precise prediction of information diffusion (a.k.a. information cascade) has been proven crucial to some valuable applications, including market forecast \cite{shen2014modeling}, community detection \cite{barbieri2013cascade}, etc. 

Inspired by the success of deep neural networks in computer vision, recommender systems, etc., more and more researchers also employed DNNs to model information cascade. In recent years, some deep models based on recurrent neural networks (RNNs) \cite{islam2018deepdiffuse,yang2018neural}, including the RNNs coupled with attention mechanism~\cite{wang2017cascade}, have been proposed to predict information diffusion, since an information cascade is often modeled as a sequence.

\begin{figure}[!htb]
\centering 
\includegraphics[width=4.2in]{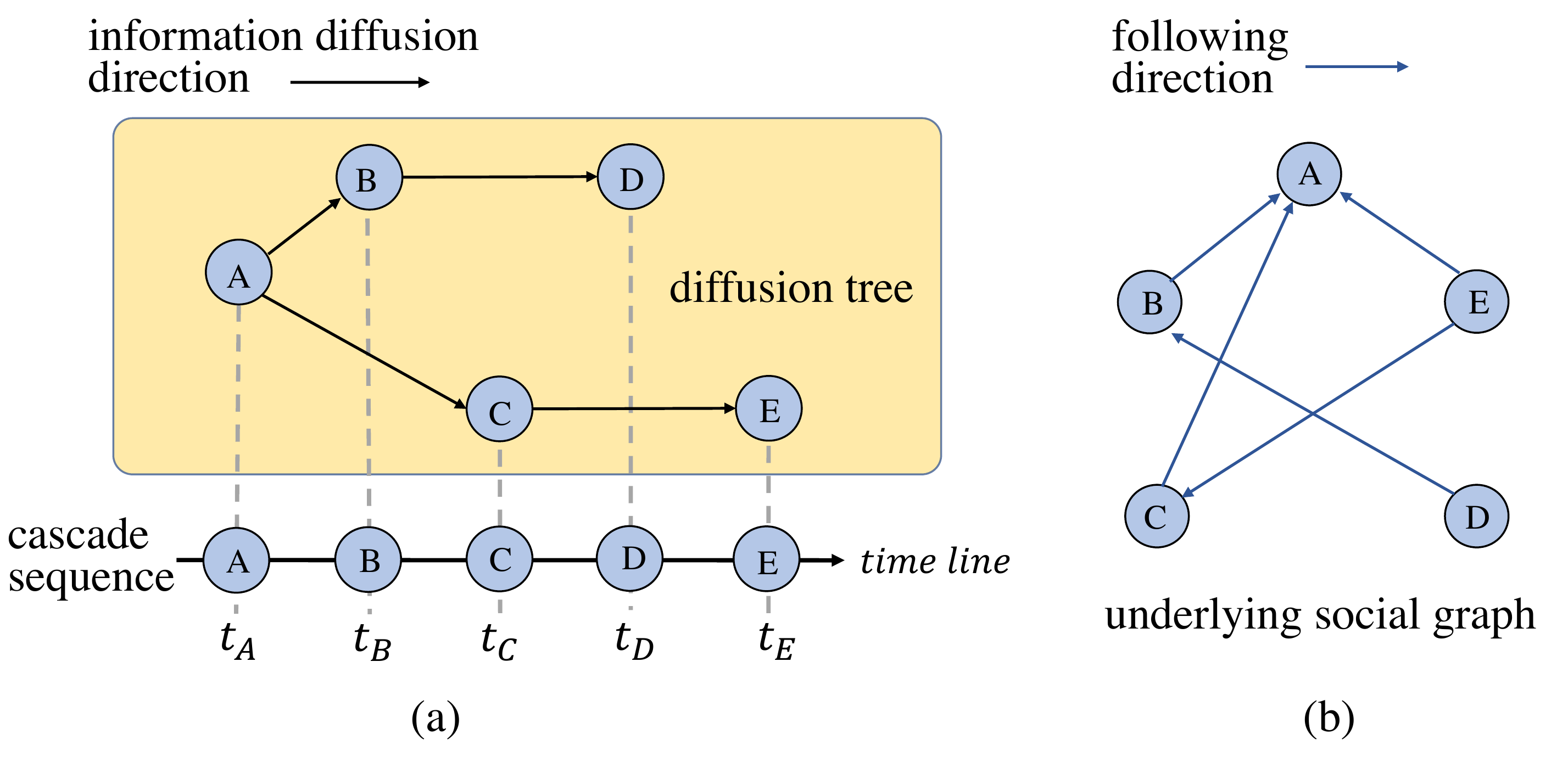}
\caption{An example of an information cascade sequence, diffusion tree and social graph.} 
\label{diffusion_tree}
\end{figure} 

Despite the effects of these RNN-based models, their sequential assumption may model the user dependencies inconsistent with the diffusion tree. We take the example shown in Fig. \ref{diffusion_tree} to elaborate it. Given the following information cascade $c=\big\{(A,t_{A}),(B,t_{B}),(C, t_{C}),(D, t_{D}),(E, t_{E})\big\}$ where user $A\sim E$ are sorted according to their activation (disseminating information) time. User $A$ is the cascade source while user $E$ is last one who disseminates this piece of information. There is a real diffusion tree behind this cascade sequence which does not strictly follow the sequential assumption. Suppose a set of chronological diffusion behaviors is denoted as $\{(A,B,t_{B}),(A,C,t_{C}),(B,D,t_{D}),(C,E,t_{E})\}$ where a triplet $(A,B,t_{B})$ means $A$ spreads information to $B$ at time $t_{B}$. Based on this diffusion behavior set, a real diffusion tree can be reconstructed as shown in the yellow rectangle of Fig. \ref{diffusion_tree}(a). The diffusion tree indicates that $D$ is influenced by $B$ rather than $C$ to be activated. The RNN-based models fed only with $c$ would model $D$ to be influenced by $C$ more than $B$, since they assume one object in a sequence depends on an immediate predecessor rather than a non-immediate predecessors. Such modeled user dependency is inconsistent with the diffusion tree, but could be alleviated if the underlying social graph of these five users (as depicted in Fig. \ref{diffusion_tree}(b)) is provided to guide information cascade modeling. It shows that the social topology plays an important role in precise cascade modeling. 

Although some sequential models \cite{chen2019information,li2017deepcas,liu2020cascade,wang2017topological,yang2019multi,wang2018sequential} have taken social topology into account when modeling cascade process, they are still not satisfactory due to their inherent sequential assumption. More recently, a non-sequential model with hierarchical attention networks ~\cite{wang2019hierarchical} has demonstrated good performance through capturing non-sequential dependencies among the users in a cascade, but this model still neglects the underlying social graph. 

In addition, the user dependencies captured by most of previous diffusion models \cite{wang2019hierarchical,yang2019multi,wang2017cascade} are only unidirectional. They suppose a successor is only influenced by a predecessor during information diffusion, whereas the opposite dependency is rarely considered. According to \emph{Laswell’s `5W' Communication Model} \cite{lasswell1948structure}: Who (sender) says What in Which channel to Whom (receiver) with What effect, we believe that a user's role in the process of an information cascade is not just a single role of sender or receiver, but a \emph{dual role of both sender and receiver}. In other words, information diffusion depends not only on how each user in a cascade behaves as a sender of his/her successors, but also on how each user behaves as a receiver of his/her predecessors. For example, a Twitter user is easy to be influenced by his/her followees who often disseminate appealing information to him/her. Meanwhile, a users may also be influenced by his/her followers when he/she decides to disseminate information, i.e., he/she would consider what kind of information the followers are more likely to receive. Accordingly, capturing such dual role user dependencies instead of single role (unidirectional) user dependencies is helpful to precise cascade prediction. Unfortunately, such intuition was overlooked by previous cascade models.

Inspired by above intuitions, in this paper we propose a non-sequential deep model of information cascade, named as {\bf TAN-DRUD} ({\bf T}opology-aware {\bf A}ttention {\bf N}etworks with {\bf D}ual {\bf R}ole {\bf U}ser {\bf D}ependency). TAN-DRUD is built with two-level attention networks to learn optimal cascade representations, which are crucial to predict information diffusion precisely. At first, the \emph{user-level attention} network is used to learn the \emph{dependency-aware representation} of a user that serves as the basis of a cascade's representation. Specifically, each user is first represented by two separate embeddings corresponding to his/her dual role of information sender and receiver, respectively. In order to exploit the social topology's indicative effects on information cascade modeling, we employ Node2Vec \cite{grover2016node2vec} to learn \emph{topological embeddings} upon the social graph, which are used to adjust the attention values. Through our empirical studies, we have verified that Node2Vec is more effective than GNN-based graph embedding model ~\cite{zhang2020sce} in our model framework. With the topology-adjusted attentions, the user dependencies among a cascade are encoded into \emph{dependency-aware user representations} dynamically and sufficiently. Then, the \emph{cascade-level attention} network is fed with the combination of dependency-aware user representations, topological embeddings and time decay information, to learn the cascade's representation. Since the cascade-level attentions can be regarded as a historical activated user's probability of activating the next user, the diffusion trees can be inferred based on our model. 
 Our contributions in this paper are summarized as follows:
\begin{enumerate}
\item We propose a non-sequential cascade model TAN-DRUD with two-level attention networks, which demonstrates satisfactory performance through capturing dual role user dependencies in information cascades. What's more, TAN-DRUD's performance is enhanced with the aid of social topology.

\item In the user-level attention network of TAN-DRUD, we particularly design a sender attention module and a receiver attention module to learn two separate embeddings for a user in a cascade, which encode dual role user dependencies sufficiently. Such manipulation’s rationale is inspired by the classic communication model, and has been proven more effective than modeling user dependencies in terms of single role. 

\item We conducted extensive experiments on three real cascade datasets, which not only justify our model's advantage over the state-of-the-art diffusion models, but also demonstrate our model's capability of inferring diffusion tree.
\end{enumerate}

The rest of this paper is organized as below. In Section 2, we introduce some research works related to information cascade modeling. Next, we formalize the problem addressed in this paper and introduce our proposed model in detail in Section 3. In Section 4, we display our experiment results based on which we provide further analysis. At last, we conclude our work in Section 5.

\section{Related Work}
Information cascade models in deep learning domain can be divided into two types: diffusion path based models and topological-based diffusion models. 
\subsection{Diffusion Path based Methods}
DeepCas \cite{li2017deepcas} is the first end-to-end deep learning model for information diffusion prediction. It uses the same way as DeepWalk \cite{perozzi2014deepwalk} to sample node sequences from cascade graphs, then bidirectional gated recurrent units (GRU) \cite{GRU} and attention mechanism are used to process node sequences and get cascade representations for prediction. DCGT \cite{li2018joint} is an extended model of DeepCas which incorporates the content of information to predict cascades. \cite{cao2017deephawkes} is an RNN-based model fed with diffusion sub-paths, in which a non-parametric time attenuation factor is applied on the last hidden state of all sub-paths, to represent the self-excitation mechanism \cite{self-exciting} in information diffusion. DeepDiffuse \cite{islam2018deepdiffuse} utilizes timestamps and cascade information to focus on specific nodes for prediction. \cite{yang2018neural} employs self-attention and CNNs to alleviate RNN's disadvantage of long-term dependency. Unlike above sequential models, HiDAN~\cite{wang2019hierarchical} is a non-RNN hierarchical attention model which captures dependency-aware user representations in a cascade, resulting in precise prediction.

\subsection{Topological-based Diffusion Model}
 TopoLSTM~\cite{wang2017topological} utilizes directed acyclic (social) graph to augment node embedding learning in a cascade. \cite{yang2015rain} studies the influence of interactions between users’ social roles during information diffusion.  A role-aware information diffusion model is proposed to combine social role information and diffusion sequences information. \cite{chen2019information} builds a shared presentation layer to discover the complementary information of sequence representations and topology representations, which can be applied to micro and macro tasks. FOREST \cite{yang2019multi} employs reinforcement learning on social graph modeling to solve multi-scale tasks for information diffusion prediction. \cite{liu2020cascade} uses multi self-attention and social graph information to discovery the long-term dependency, which is crucial for diffusion prediction. Inf-vae \cite{sankar2020inf} learns topological embeddings through a VAE framework as encoders and decoders, then an co-attentive fusion network is used to capture complex correlations between topological and temporal embeddings to model their joint effects. \cite{wang2020joint} discovers relations between information diffusion and social network through an identity-specific latent embeddings. DyHGCN \cite{yuan2020dyhgcn} implements a dynamic heterogeneous graph convolutional network to capture users' dynamic preferences between global social graph and temporal cascade graph.
\section{Methodology}

\subsection{Problem Definition}
Suppose an observed cascade sequence consists of $i$ users along with their timestamps of information dissemination. This sequence is denoted as $$c_i= \{(u_{1},t_{1}), (u_{2},t_{2}),\ldots, (u_{i},t_{i})\}$$ where $u_j (1\leq j \leq i)$ is a user and $t_j$ is $u_j$'s timestamp. All users are sorted chronologically, i.e., $t_1<t_2<...<t_i$. Moreover, an underlying social graph $\mathcal{G=(V,E)}$ including all users may be obtained, where each user corresponds to a node in set $\mathcal{V}$ and $|\mathcal{V}|=N$. Then, the prediction task of our model is formalized as: given $c_i$, the model should predict the next user to be activated, denoted as $u_{i +1}$, through computing the conditional probability $p(u_{i+1} | c_i)$.

\begin{figure*}[!htb]
\centering 
\includegraphics[width=\linewidth]{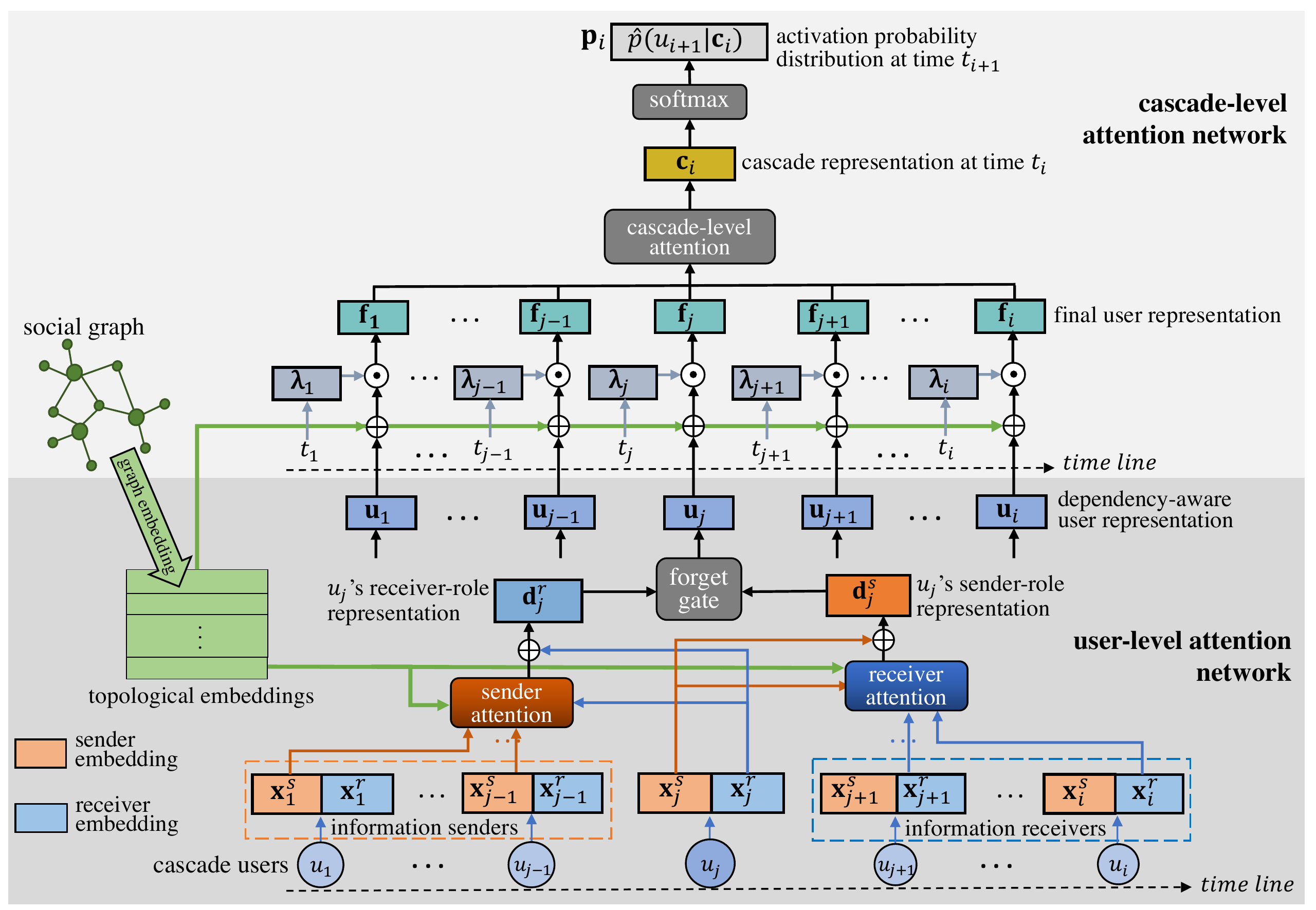}
\caption{The framework of our proposed TAN-DRUD consisting of two-level attention networks. In the user-level attention network, each user in a cascade is first represented by a sender embedding and a receiver embedding. With these two embeddings and the topological embeddings, the user's dependency-aware representation is learned. In the cascade-level attention network, a cascade's representation is generated based on dependency-aware user representations, topological embeddings and time decay, with which the next activated user is predicted. } 
\label{framework}
\end{figure*}

\subsection{Model Framework}
The framework of our model is depicted in Fig.\ref{framework}. In the user-level attention network, each user is first represented by a \emph{sender embedding} and a \emph{receiver embedding} corresponding to his/her dual role. Then, a user's \emph{receiver-role representation} is learned as the attentive sum of his/her predecessors' sender embeddings. Likewise, a user's \emph{sender-role representation} is learned as the attentive sum of his/her successors' receiver embeddings. Furthermore, we utilize the social \emph{topological embeddings} learned by Node2Vec to adjust the attention values. Next, a forget gate mechanism is used to aggregate a user's dual role representation into the user's \emph{dependency-aware representation}. In cascade-level attention network, dependency-aware user representations, social topological embedding, and time information are combined to generate a cascade's representation for diffusion prediction.

\subsection{Embedding Preparation}
\subsubsection{Sender and Receiver Embeddings}
Two separate embeddings are first initialized to represent a given user. We use $\mathbf{X}^s, \mathbf{X}^r \in \mathbf{R}^{N \times d}$ to represent sender embedding matrix and receiver embedding matrix respectively. Each row of the two matrices represents a user. $N$ is the total number of users, and $d$ is embedding dimension (size). All embeddings in $\mathbf{X}^s, \mathbf{X}^r$ are initialized in random.

\subsubsection{Social Topological Embeddings}
Since $\mathcal{G}$ is a homogeneous network, the models towards heterogeneous networks such as Metapath2Vec \cite{M2V} are not suitable for our scenario. In addition, we do not prefer the semi-supervised graph neural networks (GNNs), e.g., GCN, because such graph embedding models were primarily designed to aggregate node features which are subject to specific downstream tasks. Through our empirical study, we finally adopted Node2Vec to learn topological embeddings.

\subsection{Two-level Attention Networks}

\subsubsection{User-level Attention Network}
Given $u_j$ in a cascade $\{(u_{1},t_1), \ldots, (u_{i},t_i)\}$, we first design a \emph{sender attention module} to learn $u_j$'s receiver-role representation $\mathbf{d}_j^r$ which is computed based on the sender embeddings of  $\{u_{1}, \ldots, u_{j-1}\}$. We symmetrically learn $u_j$'s sender-role representation $\mathbf{d}_j^s$, which is computed based on the receiver embeddings of $\{u_{j+1}, \ldots, u_{i}\}$ in \emph{receiver attention module} .  

Formally, suppose $u_k (1\leq k \leq j-1)$ is a predecessor of $u_j$, $u_k$'s sender attention to $u_j$ is $\alpha_{kj}^s$ and computed as
\begin{equation}\label{eq:alpha1}
{\alpha'}_{k j}^s=\frac{\exp \left(\left\langle\mathbf{W}_{s}^{o} \mathbf{x}_{k}^s, \mathbf{W}_{r}^{c} \mathbf{x}_{j}^r\right\rangle\right)}{\sum_{l=1}^{j-1} \exp \left(\left\langle\mathbf{W}_{s}^{o} \mathbf{x}_{l}^s, \mathbf{W}_{r}^{c} \mathbf{x}_{j}^r\right\rangle\right)},
\end{equation}
\begin{equation}\label{eq:E}
\mathbf{E} = \frac{\mathbf{G}_{c} \mathbf{G}_{c}^\top}{||\mathbf{G}_{c}|| \times ||\mathbf{G}_{c}^\top|| }, 
\end{equation}
\begin{equation}\label{eq:alpha2}
\alpha_{k j}^s = \frac{\exp{({\alpha'}_{k j}^s e_{k j}) }}{\sum_{l=1}^{j-1} \exp{({\alpha'}_{l j}^s e_{l j})}}.
\end{equation}
In Eq. \ref{eq:alpha1}, $\mathbf{x}_{k}^s$ is $u_k$'s sender embedding obtained from $\mathbf{X}^s$ and $\mathbf{x}_{j}^r$ is $u_j$'s receiver embedding obtained from $\mathbf{X}^r$. $\mathbf{W}_{s}^{o}, \mathbf{W}_{r}^{c} \in \mathbf{R}^{d\times d}$ are transformation matrices and $<,>$ is inner product. According to Eq. \ref{eq:alpha1}, the attention ${\alpha'}_{k j}^s$ captures $u_j$'s dependency on $u_k$ in terms of the information receiver of $u_k$. In Eq. \ref{eq:E}, $\mathbf{G}_c\in \mathbf{R}^{i\times d_g}$ is the matrix consisting of the topological embeddings of $\{u_{1}, \ldots, u_{i-1}, u_{i}\}$. Thus, $\mathbf{E}$ quantifies the social similarities between $u_1\sim u_i$. Two social similar users are more likely to influence each other during information diffusion.  Hence we utilize $\mathbf{E}$'s element $e_{kj}, e_{lj}$ to adjust ${\alpha'}_{k j}^s$ into $\alpha_{k j}^s$ as Eq. \ref{eq:alpha2}. Then $u_j$'s receiver-role representation $\mathbf{d}_j^r$ is computed as follows, 

\begin{equation}
{\mathbf{d}'}_{j}^r=\sum_{k=1}^{j-1} \alpha_{k j}^s \mathbf{x}_{k}^s,\quad \mathbf{d}_{j}^r={\mathbf{d}'}_{j}^r+\mathbf{x}_{j}^r.
\end{equation}

Similarly, $u_j$'s sender-role representation $\mathbf{d}_{j}^s$ is computed as
\begin{equation}
{\alpha'}_{k j}^r=\frac{\exp \left(\left\langle\mathbf{W}_{r}^{o} \mathbf{x}_{k}^r, \mathbf{W}_{s}^{c} \mathbf{x}_{j}^s\right\rangle\right)}{\sum_{l=j+1}^{i} \exp \left(\left\langle\mathbf{W}_{r}^{o} \mathbf{x}_{l}^r, \mathbf{W}_{s}^{c} \mathbf{x}_{j}^s\right\rangle\right)},
\end{equation}
\begin{equation}\label{eq:alpha3}
\alpha_{k j}^r = \frac{\exp{({\alpha'}_{k j}^r e_{k j}) }}{\sum_{l=j+1}^{i} \exp{({\alpha'}_{l j}^r e_{l j})}},
\end{equation}
\begin{equation}
{\mathbf{d}'}_{j}^s=\sum_{k=j+1}^{i} \alpha_{k j}^r \mathbf{x}_{k}^r, \quad \mathbf{d}_{j}^s={\mathbf{d}'}_{j}^s+\mathbf{x}_{j}^s.
\end{equation}

Next, we need to aggregate $u_j$'s sender-role representation and receiver-role representation into one dependency-aware representation $\mathbf{u}_j$. We use a forget gating mechanism ~\cite{wang2017topological} to implement this operation as follows, since it can wisely decide how much input information should be reserved or forgotten to compose the output.

\begin{equation}
\mathbf{m}=\sigma(\mathbf{W}_{m}^{s} \mathbf{d}_{j}^s+\mathbf{W}_{m}^{r} \mathbf{d}_{j}^r+\mathbf{b}_{m}),
\end{equation} 
\begin{equation}
\mathbf{n}=\sigma(\mathbf{W}_{n}^{s} \mathbf{d}_{j}^s+\mathbf{W}_{n}^{r} \mathbf{d}_{j}^r+\mathbf{b}_{n}),
\end{equation} 
\begin{equation}
\mathbf{u}_j=(\mathbf{1}-\mathbf{m}) \odot \mathbf{d}_{j}^s + (\mathbf{1}-\mathbf{n}) \odot \mathbf{d}_{j}^r
\end{equation}
where $\sigma$ is Sigmoid function and $\odot$ is element-wise product. $\mathbf{1}\in \mathbf{R}^d$ is a unit vector. All $\mathbf{W}$s $\in \mathbf{R}^{d \times d}$ and $\mathbf{b}$s $\in \mathbf{R}^{d}$ in above equations are trainable parameters. $\mathbf{d}_{j}^s$ and $\mathbf{d}_{j}^r$  generated by the user level attention mechanism are input into this gated model to obtain the user-level representation of the user. 
Specifically, vector $\mathbf{m}, \mathbf{n}\in\mathbf{R}^d$ are used to control how much information we should forget
for each type of inputs. 

\subsubsection{Cascade-level Attention Network}
Three kinds of information are used in this step: dependency-aware user representation, topological information and time decay.  We first map all topological embeddings in $\mathbf{G}_c$ into the embeddings of $d$ dimensions as follows:
\begin{equation}\label{eq:G}
\mathbf{G}_{new}= \operatorname{tanh} \left(\mathbf{W}_{g} \mathbf{G}_c+\mathbf{b}_{g}\right)
\end{equation}where $\mathbf{G}_{new}\in \mathbf{R}^{i \times d}$ is transformed topological embedding matrix. $\mathbf{W}_{g} \in \mathbf{R}^{d \times d_g}$ and $\mathbf{b}_{g} \in \mathbf{R}^{d}$ are trainable parameters. 

To consider time decay, we first set a unit of time interval as $\Delta^t=T_{\max }/T$, where $T_{max}$ is the max time interval observed from all cascades, and $T$ is the number of time intervals. Then, given $u_j$'s time decay interval $\Delta t_{j}=t_{i}-t_{j}$, we get its time decay vector $\mathbf{t}_{j} \in \mathbb{R}^{T}$ (one-hot vector), in which only the $n$-th element is 1 when $n = int(\Delta t_{j}/ \Delta^t)$. $int(\cdot)$ is rounding up operation. We also map $\mathbf{t}_j$ into an embedding with the same size as user representations by:
\begin{equation}\label{eq:t}
\boldsymbol{\lambda}_{j}=\sigma(\mathbf{W}_{t} \mathbf{t}_{j}+\mathbf{b}_{t})
\end{equation}
where $\mathbf{W}_{t} \in \mathbb{R}^{d \times T}$ and $\mathbf{b}_{t} \in \mathbb{R}^{d}$ are trainable parameters. 

Then, $u_j$'s final representations is computed as
\begin{equation}\label{eq:F}
    \mathbf{f}_j = \boldsymbol{\lambda}_{j} \odot (\mathbf{g}_j + \mathbf{u}_j)
\end{equation}where $\mathbf{g}_j$ is $u_j$'s topological embedding obtained from $\mathbf{G}_{new}$ in Eq. \ref{eq:G}.

Next, $u_j$'s time-aware influence to the next activated user can be quantified by the following attention:
\begin{equation}\label{eq:beta}
    \beta_{j}=\frac{\exp \left(\left\langle\mathbf{w}, \mathbf{f}_{j}\right\rangle\right)}{\sum_{k=1}^{i} \exp \left(\left\langle\mathbf{w},  \mathbf{f}_{k}\right\rangle\right)}
\end{equation}
where $\mathbf{w} \in \mathbf{R}^{d}$ is a trainable embedding. 

At last, for a cascade observed until time $t_i$, i.e., $c_i$, its representation $\mathbf{c}_i$ is computed based on the adjusted representations of the users in $c_i$. Thus we get
\begin{equation}
    \mathbf{c}_{i}=\sum_{j=1}^{i} \beta_{j} \mathbf{f}_{j}.
\end{equation}

\subsection{Prediction and Optimization}
Given $\mathbf{c}_{i}$, the activation probability distribution of all users at time $t_{i+1}$ is denoted as $\mathbf{p}_i\in \mathbb{R}^N$, and computed as:
\begin{equation}\label{eq:p}
    \mathbf{p}_i=\operatorname{softmax}(\mathbf{W}_{c} \mathbf{c}_{i}+\mathbf{b}_{c})
\end{equation}where $\mathbf{W}_{c} \in \mathbf{N}^{d \times T}$ and $\mathbf{b}_{c} \in \mathbf{R}^{N}$ are trainable parameters.

Given the training set containing $M$ cascade sequences in which the $m$-th cascade is denoted as $c^m$ and its length is $n_m$, TAN-DRUD's learning objective is to minimize the following log-likelihood loss:
\begin{equation}
    \mathcal{L}=-\frac{1}{M}\sum_{m=1}^{M} \sum_{i=1}^{n_{m}-1} \log \hat{p}(u_{i+1} \mid \mathbf{c}_{i}^{m}) + \lambda L_{2}
\end{equation}
where $u_{i+1}$ is the truly activated user at time $t_{i+1}$ given $c^m$, and $\hat{p}(u_{i+1} \mid \mathbf{c}_{i}^{m})$ is fetched from the $\mathbf{p}_i$ computed according to Eq. \ref{eq:p}. $\lambda$ is the controlling parameter of L2 regularization. We use stochastic gradient descent and Adam algorithm for optimization.

\section{Experiments}
In this section, we try to answer the following research questions through our empirical studies.

{\bf RQ1}: Is our TAN-DRUD more effective and efficient than the state-of-the-art diffusion models?

{\bf RQ2}: Are the two separate embeddings corresponding to a user's dual role helpful for enhanced prediction performance?

{\bf RQ3}: Is the incorporated social topology helpful for prediction performance?

{\bf RQ4}: Is the graph embedding model sensitive to TAN-DRUD's final performance?

{\bf RQ5}: Can the diffusion trees be recovered approximately by our model?

\subsection{Datasets and Baselines}
We conducted experiments upon the following three datasets often used in information diffusion prediction to evaluate our model.

\textbf{Twitter}~\cite{hodas2014simple}: As the most prevalent social media, tweet spreading among Twitter users is a representative kind of information diffusion in social networks. This dataset contains the tweets with URLs posted in Oct, 2010. The tweets with the same URL are regarded as an information cascade, thus their publishers are the activated users in this cascade.

\textbf{Douban}~\cite{zhong2012comsoc}: It is a review sharing website for various resources including book, movie, music, etc., where users seek their favorite resources based on others' reviews. In this dataset, each book is regarded as a piece of information. A user is regarded as being activated and joining the cascade of a book if he/she has read this book. Similar to Twitter, each user in Douban can also follow others.

\textbf{MemeTracker}~\cite{leskovec2009meme}:  It collects massive news stories and blogs from online websites and tracks popular quotes and phrases as memes. We treat each meme as a piece of information and each website URL as an activated user. Thus, social topology does not exist in this dataset, which is used to evaluate the models without topological information.

The detailed statistics of the datasets are shown in Table \ref{tbl1}. We divided all cascades in each datasets into training set, validation set and test set, according to the ratio of 8:1:1. TAN-DRUD's source codes and our experiment samples on \url{https://github.com/JUNJIN0126/TAN-DRUD}

\begin{table}[!htb]
\caption{Statistics of the three used datasets.}\label{tbl1}
\centering
\begin{tabular}{|M{120pt}|M{45pt}|M{45pt}|M{45pt}|}
\bottomrule 
\textbf{Dataset} & \textbf{Twitter} & \textbf{Douban} & \textbf{Meme} \\
\hline
 User number & 12,627 & 23,123 & 4,709 \\
 Cascade number & 3,442 & 10,602 & 12,661 \\
 Average cascade length & 32.60 & 27.14 & 16.24 \\
 Social link number & 309,631 & 348,280  & --  \\
\toprule
\end{tabular}
\end{table}

We compared TAN-DRUD with the following cascade models, including sequential models and non-sequential models to verify our TAN-DRUD's advantages.

\indent\textbf{DeepDiffuse}~\cite{islam2018deepdiffuse}: It employs embedding technique on timestamps and incorporates cascade information to focus on specific nodes for prediction. 

\textbf{Bi-LSTM}~\cite{huang2015bidirectional}: The dual role user dependencies can also be regarded as bi-directional user dependencies in a sequence. So Bi-LSTM consisting of forward LSTM and backward LSTM can be used to model cascade sequences, and then predict the next activated user. 

\textbf{TopoLSTM}~\cite{wang2017topological}: It is an LSTM-based model incorporated with diffusion topology, i.e., directed acyclic graph for diffusion prediction.

\textbf{SNIDSA}~\cite{wang2018sequential}: It uses attention networks to extract user dependencies from social graph, then adopts a gate mechanism to combine user information and sequential information. 

\textbf{FOREST} ~\cite{yang2019multi}: It employs reinforcement learning framework fed with social graphs to solve multi-scale tasks for information diffusion.


\textbf{HiDAN} ~\cite{wang2019hierarchical}: It is a non-sequential model built with hierarchical attention networks. Compared with TAN-DRUD, it does not establish two separate embeddings for users, and omits social graphs.

In addition, we propose a variant \textbf{AN-DRUD} of TAN-DRUD for ablation study, in which social topological embeddings are absent. 

\subsection{Experiment Settings}
We introduce some important settings of our experiments as follows.
\subsubsection{Evaluation Metrics}
The next infected user prediction can be regarded as a ranking problem based on users' potential probabilities of spreading the information. Thus, we adopted \textit{Precision on top-K ranking} (P@K) and \textit{Reciprocal Rank} (RR) as our evaluation metrics, since they are popular to evaluate sequential ranking~\cite{wang2019hierarchical,wang2017cascade}. Specifically, given a test sample, its P@K=100\% if the true next activated user $u_{i+1}$ is in its top-K list ranked according to $\hat{p}$, otherwise P@K=0.

\subsubsection{General Settings}
We ran the experiments on a workstation with GPU of GeForce GTX 1080 Ti. For the baseline diffusion models and graph embedding models, we directly used their public source codes, and tuned their hyper-parameters in terms of optimal prediction performance. The topological embedding size ($d_g$) was set to 128.

\subsubsection{TAN-DRUD's Settings}
TAN-DRUD's hyper parameters includes learning rate, topological embedding size, user dual role embedding size, time interval number and so on. We set learning rate to 0.001, $\lambda$=1e-5, and also used Dropout with the keep probability of 0.8 to enhance our model's generalization capability. Due to space limitation, we only display the results of tuning user dual role embedding dimension $d$ and time interval number $T$. 

In general, the embedding dimension (size) in deep learning models is set empirically. User information in cascades will not be captured sufficiently if user dual-role embedding size is too small, while overfitting and high training time consumption may happen if it is too large. Therefore, we set $d$ to different values to investigate its influence on our model’s performance. According to the results in in Table \ref{tbln}, we set $d=64$ for our model in the subsequent comparison experiments.
\begin{table}[!htb]
\centering
\caption{TAN-DRUD's prediction performance (score \%)  with dual role embedding sizes ($d$).}\label{tbln}
\begin{tabular}{|M{40pt}|M{40pt}|M{40pt}|M{40pt}|M{40pt}|}
\hline
\textbf{$d$} & \textbf{RR} & \textbf{P@10} & \textbf{P@50} & \textbf{P@100} \\
\hline 
16 & $14.31$ & $24.33$ & $43.56$ & $53.50$ \\
32 & $15.53$ & $26.30$ & $45.12$ & $54.58$ \\
\textbf{64} & \textbf{16.62}& \textbf{28.13} & \textbf{45.61} & \textbf{55.43} \\
128 & $15.82$ & $27.53$ & $45.17$ & $54.62$ \\
\hline
\end{tabular}
\end{table}

The time interval number $T$ is used to generate the time decay vector in Eq. \ref{eq:t}. Small $T$ results in coarse-grained representations of time decay between different users in a cascade. Thus the temporal features can not contribute to precise cascade modeling. By contrast, large $T$ leads to the time decay vector of large size, resulting in high training time consumption. According to the results in in Table \ref{tblm}, we set $T=50$ for our model in the subsequent comparison experiments.

\begin{table}[!htb]
\centering
\caption{TAN-DRUD's prediction performance (score \%) with different time interval numbers ($T$).}\label{tblm}
\begin{tabular}{|M{40pt}|M{40pt}|M{40pt}|M{40pt}|M{40pt}|}
\hline
\textbf{$T$} & \textbf{RR} & \textbf{P@10} & \textbf{P@50} & \textbf{P@100} \\
\hline 
1 & $15.95$ & $27.12$ & $45.53$ & \textbf{55.91} \\
10 & $15.84$ & $27.26$ & $45.06$ & $55.04$ \\
\textbf{50} & \textbf{16.62}& \textbf{28.13} & 45.61 & 55.43 \\
100 & $16.32$ & $27.52$ & \textbf{46.04} & $55.36$ \\
\hline
\end{tabular}
\end{table}

\begin{table*}[t]
\caption{Prediction performance (score \%) of all compared models for the three datasets. The best performance scores among all compared models are indicated in bold. The performance scores of leading baseline are underlined.}\label{tbl2}
\centering
\begin{tabular}
{|p{65pt}|p{24pt}|p{22pt}|p{22pt}|p{26pt}|p{22pt}|p{22pt}|p{22pt}|p{26pt}|p{24pt}|p{22pt}|p{22pt}|p{26pt}|}
\bottomrule 
\multirow{2}{*} {\diagbox{Model}{Dataset}}& \multicolumn{4}{|c|} { \textbf{Twitter} } & \multicolumn{4}{c|} { \textbf{Douban} } & \multicolumn{4}{c|} { \textbf{Meme} } \\
\cline{2-13}
  & RR & P@10 & P50 & P100 & RR & P@10 & P@50 & P@100 & RR & P@10 & P@50 & P@100 \\
 \hline
 DeepDiffuse & 2.21& 4.45 & 14.35 & 21.61 & 3.23 & 9.02 & 14.93 & 19.13 & 6.48 & 13.45 & 30.10 & 41.31  \\
Bi-LSTM & 7.12 & 13.41 & 26.71 & 36.06 & 7.95 & 15.97 & 29.89 & 37.41 & 12.32 & 24.73 & 46.27 & 56.33 \\
Topo-LSTM & 4.56 & 10.17 & 21.37 & 29.29 & 3.87 & 8.24 & 16.61 & 23.09 & -- & -- & -- & -- \\
SNIDSA & -- & 23.37  & 35.46 & 43.39 & -- & 11.81 & 21.91 & 28.37 & -- & -- & -- & -- \\
FOREST & \underline{\textbf{17.49}} & \underline{24.63} & \underline{37.73}& \underline{46.20} & 8.19 & 13.58 &23.47 & 29.69 & \underline{\textbf{16.76}} & 28.49 & 45.85 & 55.19 \\
 HiDAN & 12.99 & 22.45 & 35.51 & 43.01 & \underline{8.78} & \underline{17.40} & \underline{32.37} & \underline{40.49} & 15.31 & \underline{29.03} & \underline{50.01} & \underline{60.07}\\
 \hline
{\bf AN-DRUD} & 13.54 & 23.28 & 36.90 & 45.28 & 8.91 & 17.72 & 32.73 & 41.01 & 16.32 & \textbf{29.48} & \textbf{51.09} & \textbf{61.33} \\
{\bf TAN-DRUD} & 16.62 & 28.13 & 45.61 & 55.43 & \textbf{9.41} & \textbf{18.21} & \textbf{34.26} & \textbf{42.02} & -- & --& -- & -- \\
improv. rate\% & -4.97 & 14.21 & 20.89 & 19.98 & 7.18 & 4.66 & 5.84 & 3.78 & -2.63 & 1.56 & 2.16 & 2.10\\
\toprule
\end{tabular}
\end{table*}

\subsection{Results and Analysis}
In this subsection, we display the results of our comparison experiments, based on which we further provide some insights.

\subsubsection{Efficacy Performance Comparison}
To answer RQ1, we first exhibit all models' mean performance scores (averaged over 5 runnings for each model) in Table \ref{tbl2}\footnote{For SNIDSA, we directly cited the results in its original paper where RR is missing.}, where Topo-LSTM, SNIDASA and TAN-DRUD's scores in Meme are absent since they need social topology. The best performance scores among all compared models are indicated in bold. The performance scores of leading baseline are underlined, which has the best performance among all baseline models. We also provide the improvement rate of our model (TAN-DRUD in Twitter and Douban, AN-DRUD in Meme) w.r.t. the leading baseline in the table's bottom row.

Based on the results, we propose the following analysis.

1. TAN-DRUD outperforms all baselines remarkably in Douban, especially in P@K. In Meme where social graphs are missing, AN-DRUD also has the best performance except for RR compared with FOREST. Please note that FOREST predicts the number of potential users in the cascade at first, and then explores the subsequent users. With a special module of \textit{cascade simulation for macroscopic prediction}, FOREST is more capable of capturing some true next activated users on first place. As a result, FOREST has the highest \textit{RR} in Twitter and Meme. However, FOREST's lower P@k implies that the rest true next activated users in it have lower ranks than that of our model. 

2. AN-DRUD's superiority over HiDAN shows that, using two separate embeddings to capture dual role user dependencies is more effective than capturing single role (unidirectional) user dependencies (to answer RQ2). AN-DRUD also outperforms Bi-LSTM, justifying that our designed mechanism is better than Bi-LSTM to capture bi-directional user dependencies in a cascade sequence. This conclusion is confirmed by the comparison results of both RNN-based models and non-sequential models, justifying that a user joins a cascade as a dual role of information sender and receiver rather than a single role.
 
3. TAN-DRUD’s superiority over AN-DRUD justifies social topology's positive effects on diffusion prediction (to answer RQ3). Moreover, TAN-DRUD’s superiority in Twitter is more prominent than Douban. It is because that Twitter is a more typical social platform where most information spreads along social links. Hence, Twitter's social topology plays a more important role in information diffusion. 

\subsubsection{Efficiency Performance Comparisons}
Then, we only compared TAN-DRUD's efficiency with the baselines which are also fed with social topology, since topology computation is generally the most time-consuming step of these models. For average time consumption of one epoch (including training and test), Topo-LSTM takes 4,915s, FOREST takes 1,800s while TAN-DRUD takes 23.7s. Even added with the time consumption of Node2Vec (695.2s), TAN-DRUD's time cost is much lower than Topo-LSTM and FOREST.  The reason of our model's higher efficiency is two-fold: 1. TAN-DRUD is a non-sequential model in which matrix computation can be parallelized. 2. We used Node2Vec to generate topological embeddings as pre-training, which avoids topology computation in every epoch.

\subsubsection{Influence of Graph Embedding Model}\label{sec:ge}
To answer RQ4, we further tested different graph embedding models' influence to TAN-DRUD's performance. In our experiments, we selected five graph embedding models suitable for homogeneous graphs, i.e., SDNE~\cite{wang2016structural}, LINE~\cite{tang2015line}, DeepWalk~\cite{perozzi2014deepwalk}, Node2Vec~\cite{grover2016node2vec} and SCE~\cite{zhang2020sce} to learn the topological embeddings fed into TAN-DRUD. The corresponding performance scores upon Twitter are shown in Table \ref{tbl3}. 
\begin{table}[!htb]
\centering
\caption{TAN-DRUD's prediction performance (score \%) upon Twitter with different graph embedding models.}\label{tbl3}
\begin{tabular}{|M{50pt}|M{40pt}|M{40pt}|M{40pt}|M{40pt}|}
\bottomrule 
\textbf{ Model} & \textbf{RR} & \textbf{P@10} & \textbf{P@50} & \textbf{P@100} \\
\hline 
SDNE & 16.27& 26.68 & 41.96 & 51.58 \\
LINE & 15.50 & 26.71 & 43.66 & 53.12  \\
DeepWalk & 16.55 & 27.84 & \textbf{46.06} & 55.25 \\
Node2Vec & \textbf{16.62} & \textbf{28.13}  & 45.60 & \textbf{55.43} \\
SCE & 13.59 & 24.20 & 40.54 & 50.75 \\
\toprule
\end{tabular}
\end{table}

On various social networks, many users may interact with or are influenced by remote users rather than their direct neighbors. Thus, the models which can capture high-order connections would be more competent to our scenario. Unlike SDNE and LINE that only model first and second order connections, DeepWalk and Node2Vec can capture high-order connections, resulting in better performance. Moreover, Node2Vec can sample more high-order neighbors than DeepWalk with breadth-first and depth-first sampling, thus it has the best performance. SCE is a new unsupervised graph embedding model based on GNN, which uses a contrastive objective based on sparsest cut problem. It may be not suitable for this relatively sparse Twitter's social graph, thus showing unsatisfactory performance compared with other graph embedding models. 

\begin{figure*}[!htb]
\centering 
\includegraphics[width=5in]{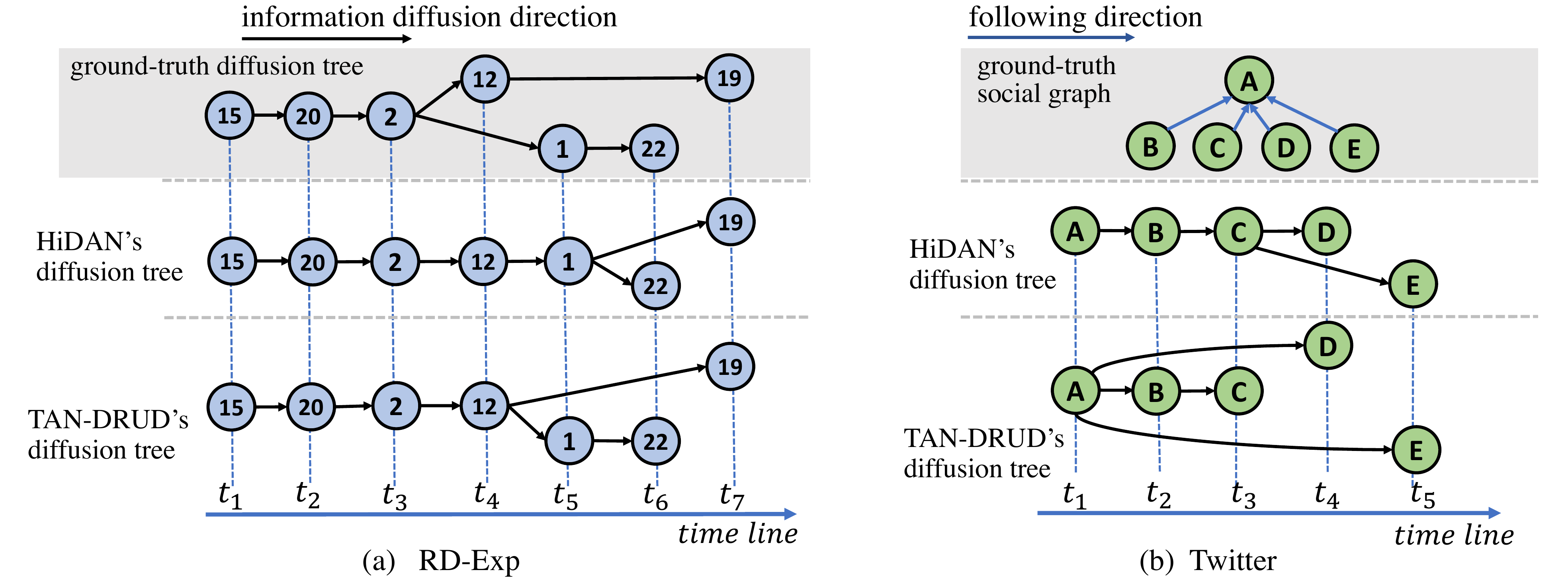}
\caption{Case study of diffusion tree inference.}\label{inferred}
\end{figure*}

\subsubsection{Diffusion Tree Inference}
In fact, the cascade-level attentions of TAN-DRUD and HiDAN quantify the probabilities of different historical activated users triggering the next user. Specifically, given a user $u_{i+1}$ activated at time $t_{i+1}$, we assume that the user in $c_i$ who has the highest attention value ($\beta_j$ in Eq. \ref{eq:beta}) triggers $u_{i+1}$, i.e., is $u_{i+1}$'s parent node in the diffusion tree. Accordingly, the diffusion tree of an observed cascade can be inferred. We illustrate two case studies to answer RQ5. The first case was extracted from the dataset RD-Exp~\cite{wang2017cascade} including social networks. This cascade sample's real diffusion tree was provided by \cite{wang2019hierarchical}. So we can compare the diffusion trees inferred by TAN-DRUD and HiDAN with the ground truth, as shown in 
Fig. \ref{inferred} (a). It shows that, TAN-DRUD's inferred diffusion tree is more approximate to the upper real diffusion tree than HiDAN's tree. In addition, Fig. \ref{inferred} (b) only displays the real social graph of a cascade's users in Twitter, since real diffusion trees do not exist in Twitter dataset. It also shows that TAN-DRUD's diffusion tree is more approximate to the real social graph, implying that TAN-DRUD's attentions are more precise than HiDAN's attentions.

\section{Conclusion}
In this paper, we propose a non-sequential information cascade model TAN-DRUD built with two-level attention networks. TAN-DRUD obtain satisfactory performance through capturing dual role user dependencies in a cascade sequence, and incorporating social topology into cascade modeling. Our extensive experiments on three real datasets demonstrate TAN-DRUD's higher efficacy and efficiency over the state-of-the-art diffusion models, and also prove that diffusion tree can be inferred approximately by our model.

\bibliographystyle{splncs03}
\bibliography{ref.bib}

\end{document}